\documentclass[aps,pra,reprint,floatfix]{revtex4-2}

\usepackage{amsmath}
\usepackage{braket}
\usepackage{amssymb}
\usepackage{graphicx}
\usepackage{xcolor}
\usepackage{siunitx}
\usepackage[normalem]{ulem}
\usepackage[hidelinks]{hyperref}
\graphicspath{{figures/}}

\newcommand{\species}[2]{\ensuremath{^{#1}\mathrm{#2}}}
\newcommand{\baf}[1]{\species{#1}{BaF}}
\newcommand{\bah}[1]{\species{#1}{BaH}}
\newcommand{\vc}{\ensuremath{v_\mathrm{cap}}}

\newcommand{\revise}[2]{{\color{#1}#2}}
\DeclareSIUnit\gauss{G}

\begin{document}

\title{\texorpdfstring{{Capture velocities for direct loading of heavy molecules into conveyor-belt magneto-optical traps}}{Capture velocities for direct loading of heavy molecules into conveyor-belt magneto-optical traps}}

\author{Shoukang Yang}
\affiliation{Zhejiang Key Laboratory of Micro-nano Quantum Chips and Quantum
	Control, School of Physics, and State Key Laboratory for Extreme Photonics
	and Instrumentation, Zhejiang University, Hangzhou 310027, China}

\author{Shuhua Deng}
\affiliation{Zhejiang Key Laboratory of Micro-nano Quantum Chips and Quantum
	Control, School of Physics, and State Key Laboratory for Extreme Photonics
	and Instrumentation, Zhejiang University, Hangzhou 310027, China}

\author{Zixuan Zeng}
\email{zixuanzeng@zju.edu.cn}
\affiliation{Zhejiang Key Laboratory of Micro-nano Quantum Chips and Quantum
	Control, School of Physics, and State Key Laboratory for Extreme Photonics
	and Instrumentation, Zhejiang University, Hangzhou 310027, China}

\author{Bo Yan}
\email{yanbohang@zju.edu.cn}
\affiliation{Zhejiang Key Laboratory of Micro-nano Quantum Chips and Quantum
	Control, School of Physics, and State Key Laboratory for Extreme Photonics
	and Instrumentation, Zhejiang University, Hangzhou 310027, China}

\begin{abstract}
	Conveyor-belt magneto-optical traps (CB-MOTs) use blue-detuned
	polarization-gradient forces to provide simultaneous cooling, confinement,
	and loading on type-II molecular transitions.
	Recent experiments with \baf{138} showed that this mechanism can directly
	load a slowed molecular beam with an efficiency exceeding that of a
	conventional red-detuned MOT.
	Here we use established optical-Bloch-equation force calculations and
	classical trajectory propagation to ask whether this direct-loading strategy
	should extend beyond the specific molecule used in the first demonstration.
	For \baf{138}, the calculation reproduces the experimentally observed trend
	that the CB-MOT capture velocity increases with laser intensity.
	We then apply the same framework to two closely related but experimentally
	distinct cases: \baf{137}, whose dense hyperfine structure complicates a
	conventional dual-frequency MOT, and \bah{138}, whose narrower linewidth and
	longer wavelength reduce the available radiative force.
	In both cases, the CB-MOT retains a broad region of nonzero capture velocity.
	These results identify the molecular conditions under which direct CB-MOT
	loading should remain effective and show that the dipole-force-dominated
	conveyor-belt mechanism provides a practical loading route for heavy
	laser-coolable molecules whose MOT performance is otherwise limited by photon
	recoil, scattering rate, or hyperfine complexity.
\end{abstract}

\maketitle

\section{Introduction}

Cold molecules are an increasingly important platform for precision
measurements, quantum simulation, quantum information processing, and controlled
chemistry \cite{
	carrColdUltracoldMolecules2009,langenQuantumStateManipulation2024,
	demilleQuantumSensingMetrology2024,cornishQuantumComputationQuantum2024,
	karmanUltracoldChemistryTestbed2024, niHighPhaseSpaceDensityGas2008,
	aikawaCoherentTransferPhotoassociated2010}.
Their rotational, vibrational, and hyperfine structure provides internal degrees
of freedom that are absent in atoms, while their permanent electric dipole
moments enable strong long-range interactions \cite{
	yanObservationDipolarSpinexchange2013,baoDipolarSpinExchangeEntanglement2023,
	hollandOnDemandEntanglementMolecules2023,vilasOpticalTweezerArray2024,
	zhangOpticalTweezerArray2022}.
The same internal structure, however, also makes molecular laser cooling and
trapping intrinsically challenging \cite{
	dirosaLasercoolingMoleculesConcept2004,shumanLaserCoolingDiatomic2010,
	andereggLaserCoolingOptically2018,andereggOpticalTweezerArray2019}.
A central experimental problem is therefore not only to cool molecules to low
temperature, but also to capture a large fraction of the incoming molecular beam
into a dense magneto-optical trap (MOT).

Most molecular MOTs operate on type-II transitions in order to close rotational
transitions
\cite{stuhlMagnetoOpticalTrapPolar2008,barryMagnetoOpticalTrappingDiatomic2014},
where dark states and polarization-gradient forces play a central role
{in molecular MOTs}
\cite{
	truppeMoleculesCooledDoppler2017,andereggRadioFrequencyMagnetoOptical2017,
	collopy3DMagnetoOpticalTrap2018,vilasMagnetoOpticalTrappingSubDoppler2022,
	zengThreeDimensionalMagnetoOpticalTrapping2024,Padilla2025AlFMOT,
	lasnerMagnetoOpticalTrappingHeavy2025,Dai2026MOT}.
In a red-detuned type-II MOT, Doppler cooling and Sisyphus heating can coexist
with the magnetic restoring force, producing traps that are hotter and less
dense than type-I atomic MOTs
\cite{
	tarbuttMagnetoOpticalTrappingForces2015,
	devlinThreeDimensionalDopplerPolarizationGradient2016,
	devlinLaserCoolingMagnetoOptical2018,jarvisBlueDetunedMagnetoOpticalTrap2018}.
Blue-detuned MOTs reverse the role of the sub-Doppler force and have been used
to produce colder and denser molecular samples
\cite{
	jarvisBlueDetunedMagnetoOpticalTrap2018,burauBlueDetunedMagnetoOpticalTrap2023,
	jorapurHighDensityLoading2024,liBlueDetunedMagnetoOpticalTrap2024}.
The recently proposed and realized conveyor-belt MOT (CB-MOT) is a particularly
useful blue-detuned configuration
\cite{
	hallasHighCompressionBlueDetunedMagnetoOptical2026,
	liConveyorBeltMagnetoOpticalTrapping2025,yuConveyorBeltMagnetoOpticalTrap2026,
	lyuTrappingCoolingMechanisms2026}.
In the ``1+2'' scheme, the two circular polarizations form two
moving standing waves with opposite velocities. The quadrupole magnetic field
then makes the molecular resonance position dependent, so that a molecule on
either side of the trap preferentially interacts with the conveyor belt moving
toward the field zero. In that moving frame, blue-detuned Sisyphus cooling slows
the molecule and transports it to the MOT center
\cite{liConveyorBeltMagnetoOpticalTrapping2025}.
More recently, a Zeeman-induced dark state mechanism was proposed to explain the
trapping force
\cite{lyuTrappingCoolingMechanisms2026}.

This CB-MOT changes the scaling of the loading force.
In a conventional red-detuned MOT, the force is closely tied to photon scattering
and is therefore limited by the maximum scattering rate and the single-photon
recoil.
For heavy molecules with narrow optical transitions,
such as BaF \cite{chenStructureBranchingRatios2016},
BaH \cite{mcnallyOpticalCyclingRadiative2020}
and BaOH \cite{bauseLaserCoolingCandidate2025}, this radiative-force scale is
small.
By contrast, the CB-MOT force is dominated by the optical dipole force
associated with blue-detuned Sisyphus cooling in moving polarization gradients.
The force can therefore increase with laser intensity over a range where the
photon scattering rate is already saturated or remains comparatively small
\cite{liConveyorBeltMagnetoOpticalTrapping2025,zengDirectLoadingBaF2026}.
This distinction was observed experimentally in \baf{138}, where direct loading
into a CB-MOT outperformed loading into a dual-frequency red MOT
\cite{zengDirectLoadingBaF2026}.

For MOT loading, the most relevant figure of merit is the capture velocity
\(\vc\) rather than the local damping coefficient at the trap center.
Molecules enter with finite velocities and traverse a finite beam diameter;
they are captured only if the force remains decelerating over the relevant range
of position and velocity.
A local MOT force that appears favorable near \(z=0\) and \(v=0\) may still give
poor loading if the damping changes sign at higher velocity, if the restoring
force is lost at larger magnetic field, or if the useful force occupies too
small a region of phase space.
The capture velocity therefore provides a direct bridge between a microscopic
force calculation and the experimentally measured loading efficiency.
It is therefore a more stringent and experimentally relevant metric than the
small-amplitude spring constant or damping coefficient alone.

In this work, we use a mature molecular laser-cooling calculation to test the
generality of direct CB-MOT loading by comparing the capture velocity with the
red MOT case.
We first benchmark the method using \baf{138}, whose level structure and
experimental CB-MOT performance are known \cite{
	chenStructureBranchingRatios2016,zengDirectLoadingBaF2026}.
We then extend the calculation to two experimentally relevant cases that probe
different limits of molecular MOT loading.
The fermionic isotope \baf{137} has a dense hyperfine spectrum, making a
conventional red-detuned MOT technically challenging \cite{
	kogelLasercooled137mathrmBaFMolecules2025}.
The molecule \bah{138} has a smaller radiative-force scale because of its
narrower linewidth and longer wavelength \cite{
	mcnallyOpticalCyclingRadiative2020}.
By comparing these two cases with $^{138}$BaF,{we identify the
conditions under which the CB-MOT remains effective and provide practical
parameter windows for extending direct CB-MOT loading to additional heavy
molecular species.}
\section{Numerical model}

We treat the molecular motion semiclassically.
The internal dynamics are calculated from optical Bloch equations (OBEs), and
the resulting time-averaged optical force is used to propagate classical
trajectories.
This approach follows earlier OBE treatments of molecular MOTs \cite{
	tarbuttMagnetoOpticalTrappingForces2015,
	devlinThreeDimensionalDopplerPolarizationGradient2016,
	devlinLaserCoolingMagnetoOptical2018,langinTheoryMolecularMagnetoOptical2023,
	xuNumericalStudyGrayMolasses2023,liConveyorBeltMagnetoOpticalTrapping2025},
including the role of type-II dark states and polarization-gradient forces,
while using the full level structure appropriate for the barium-containing
species considered here.

The density matrix obeys
\begin{equation}
	\dot{\rho}
	=
	- \mathrm{i} [H, \rho]
	+ \sum_{j} \left(C_{j} \rho C_{j}^{\dagger}
	- \frac{1}{2}\left\{C_{j}^{\dagger} C_{j}, \rho\right\}\right),
	\label{eq:obe}
\end{equation}
where \(C_j=\sqrt{\Gamma_j}L_j\), and \(L_j=\ket{\mathrm{g}_m}\bra{\mathrm{e}_n}\)
describes spontaneous decay from excited state \(n\) to ground state \(m\).
The index \(j\) labels all allowed decay channels, with branching rates \(\Gamma_j\).
Throughout the calculation, frequencies are expressed in angular-frequency units
and energies in units with \(\hbar=1\).

At position \(\boldsymbol{r}\), the Hamiltonian is
\begin{equation}
	\begin{aligned}
		H(\boldsymbol{r}, t)
		= & H_{0} - \boldsymbol{\mu} \cdot \boldsymbol{B}(\boldsymbol{r}) \\
		- & \boldsymbol{d} \cdot
		\sum_{\boldsymbol{k},i}
		\left[
			\hat{\varepsilon}_{\boldsymbol{k}, i}
			\mathcal{E}_{\boldsymbol{k}, i}
			\mathrm{e}^{
				\mathrm{i} \left( \boldsymbol{k} \cdot \boldsymbol{r}
				- \omega_{i} t\right)
			}
			+ \mathrm{c.c.}
			\right].
	\end{aligned}
	\label{eq:hamiltonian}
\end{equation}
Here \(H_{0}\) contains the field-free molecular structure,
\(\boldsymbol{\mu}\) and \(\boldsymbol{d}\) are the magnetic and electric dipole
operators, and \(\boldsymbol{B}(\boldsymbol{r})\) is the quadrupole magnetic
field.
The laser component \(i\) in beam direction \(\boldsymbol{k}\) has angular
frequency \(\omega_i\), polarization \(\hat{\varepsilon}_{\boldsymbol{k},i}\),
and electric-field amplitude \(\mathcal{E}_{\boldsymbol{k},i}\), with intensity
\( I=2\epsilon_{0} c \mathcal{E}^2 \).
We use the spherical-basis convention \(\boldsymbol{d}\cdot\hat{\varepsilon}=\sum_{q=-1}^{1}(-1)^{q}
d_q\varepsilon_{-q}\).

Dipole matrix elements are evaluated in the molecular eigenbasis.
For a basis state \(\ket{\eta,F,m}\), where \(\eta\) denotes all remaining
quantum numbers, the matrix elements are written as
\begin{widetext}
	\begin{align}
		\bra{\eta, F, m} d_{q} \ket{\eta', F', m'}
		= & (-1)^{F-m}\sqrt{2F+1}
		\begin{pmatrix}
			F  & 1 & F' \\
			-m & q & m'
		\end{pmatrix}
		\langle \eta,F\|\boldsymbol{d}\|\eta',F'\rangle ,
		\label{eq:dipole}         \\
		\bra{\eta,F,m} \mu_q \ket{\eta',F',m'}
		= &
		(-1)^{F-m+1}g_F\mu_B
		\sqrt{F(F+1)(2F+1)}
		\nonumber                 \\
		  & \times
		\begin{pmatrix}
			F  & 1 & F' \\
			-m & q & m'
		\end{pmatrix}
		\delta_{\eta,\eta'}\delta_{F,F'} .
		\label{eq:magnetic}
	\end{align}
\end{widetext}
The reduced matrix elements and effective Landé g-factors are chosen for each
molecular species and optical cycling transition.
For BaF, the molecular constants and branching structure are taken from Refs~\cite{
	chenStructureBranchingRatios2016,buRotationalHyperfineStructure2017,
	zhangLaserCoolingSlowing2022}.
The saturation parameter is defined as \(s=I/I_s=2\Omega^2/\Gamma^2\), with
saturation intensity \(I_s=\pi h c\Gamma/(3\lambda^3)\).

A rotating-frame transformation removes the optical carrier frequencies.
The remaining explicit time dependence arises from relative frequency offsets
between the laser components and from the phases
\(\boldsymbol{k} \cdot \boldsymbol{r}(t) - \Delta_{i} t\).
For a fixed force evaluation, the molecule is assumed to move at constant velocity,
\(\boldsymbol{r}(t) = \boldsymbol{r}_{0} + \boldsymbol{v}t\).
This approximation is valid because the optical pumping and phase-averaging
times are short compared with the motional time over which the velocity changes
appreciably.
The Doppler shift is therefore included through the time-dependent laser phases,
while the Zeeman shift is evaluated at the instantaneous position.

\begin{figure*}[tbp]
	\centering
	\includegraphics[width=0.98\textwidth]{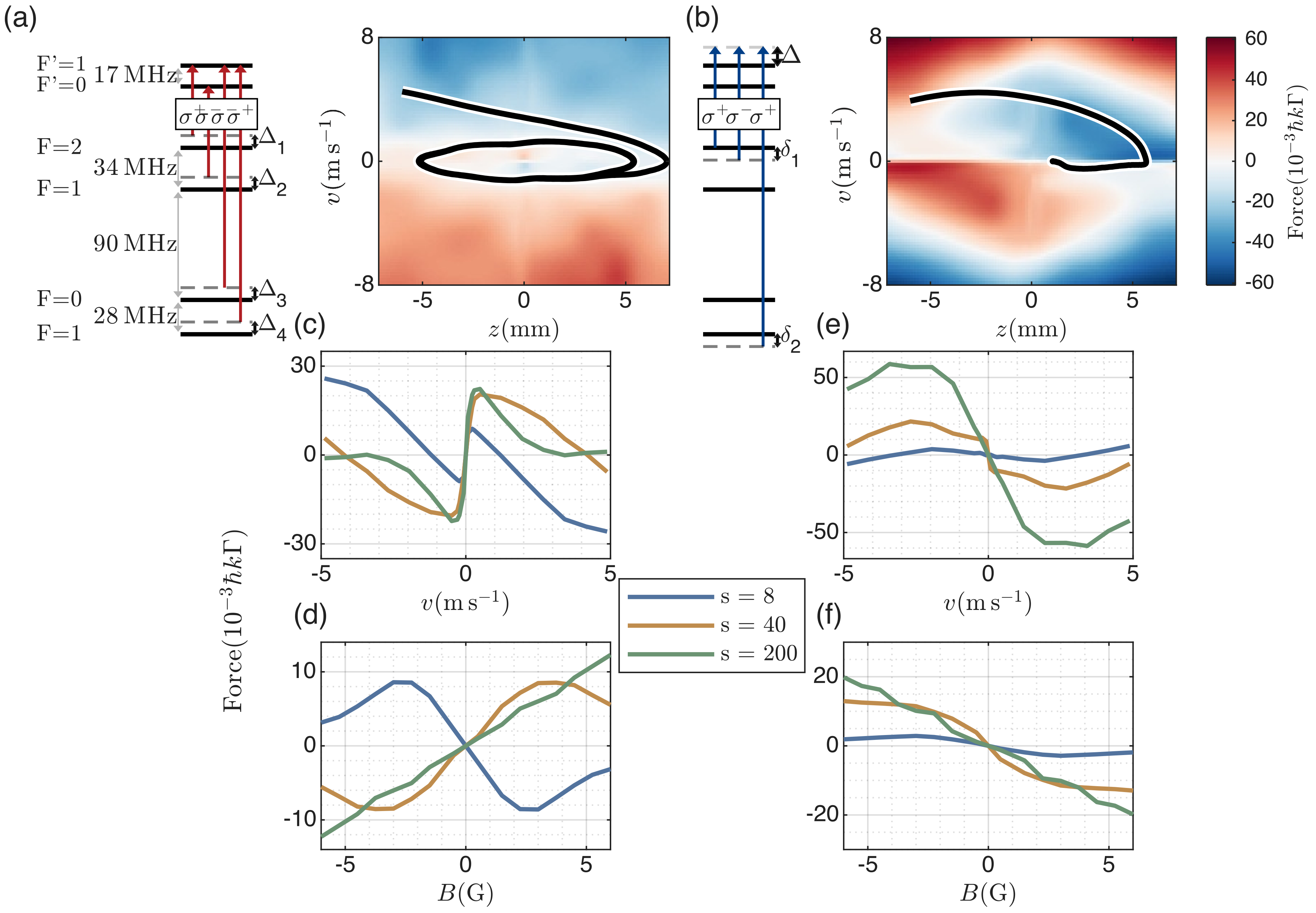}
	\caption{MOT configurations and calculated axial force maps for \baf{138}.
		(a) Dual-frequency red-detuned MOT scheme and the force map with capture
		trajectory for the red MOT with detunings
		\(\left[\Delta_1,\Delta_2,\Delta_3,\Delta_4\right]
		=2 \pi \times [-3,-14,-5,-7]\,\si{\mega\hertz}\) and saturation parameter
		\(s=16\) for each frequency component in each direction.
		(b) Three-frequency CB-MOT scheme and the force map with capture
		trajectory for the CB-MOT with single-photon detuning
		\(\Delta=2\pi \times \SI{10}{\mega\hertz}\), two-photon detunings
		\(\left[\delta_1,\delta_2\right]=2\pi \times[-2,-3]\,\si{\mega\hertz}\),
		\(s=100\)
		for each component in each direction and the magnetic gradient
		$b=12$G$/$cm. The laser beam is set to have a size of 14~mm in diameter.
		The overlaid trajectories show the dynamics of a molecule in the MOT
		with an initial velocity of 6~m$/$s.
		(c) and (d) show the calculated forces for the dual-frequency MOT at
		three laser intensities: the velocity-dependent forces at
		\(\lvert B\rvert=\SI{0}{\gauss}\) in (c), and the
		magnetic-field-dependent force at 2~m$/$s in
		(d).
		(e) and (f) show the corresponding CB-MOT forces under the same
		conditions.
	}
	\label{fig:baf138_force_map}
\end{figure*}

The optical force operator is
\begin{equation}
	\hat{\boldsymbol{F}}(t)=-\nabla H(t),
\end{equation}
and the force used for trajectory propagation is the quasi-steady time average
\begin{equation}
	\boldsymbol{f}(\boldsymbol{r},\boldsymbol{v})
	=
	\frac{1}{T}
	\int_{t_0}^{t_0+T}
	\mathrm{Tr}\!\left[
		\rho(t)\hat{\boldsymbol{F}}(t)
		\right]\,\mathrm{d}t .
	\label{eq:force_average}
\end{equation}

The averaging interval \(T\) is defined as the common period associated with the
relative laser-frequency components.
To obtain a finite \(T\), frequency components above a cutoff frequency are
neglected, and the remaining frequency differences are rounded to integer
multiples of a minimum frequency spacing.
We average the result over random initial optical phases along each axis to
remove the dependence on the molecular position within an optical wavelength.

The principal quantity calculated in this paper is the axial force map
\(F(z,v)\), obtained for a molecule moving along the MOT axis with position
\(z\) and axial velocity \(v\).
Small transverse velocities are included when estimating the capture velocity in
order to account for the finite divergence of the incident molecular beam.
For some plots we use the equivalent coordinate \(B=bz\), where \(b\) is the
axial magnetic-field gradient.
Velocity cuts are calculated at the trap center, and restoring-force cuts are
calculated by averaging the axial force over small velocities with different
directions.

Finally, the capture velocity \(\vc\) is determined by propagating molecules
through the calculated force map.
For a chosen MOT beam radius and magnetic-field gradient, a molecule is counted
as captured if it is decelerated before leaving the capture region and
subsequently remains bound near the trap center.
The reported \(\vc\) is the largest incident speed that satisfies this criterion.
This definition intentionally incorporates both velocity damping and
spatial confinement, and is therefore more restrictive than identifying the
largest velocity at which the local force is negative.

\section{Results}
We organize the results around three increasingly general questions.
The \baf{138} calculation benchmarks the model against an experimentally
realized direct-loading CB-MOT.
The \baf{137} calculation tests whether the same scheme can tolerate a denser
hyperfine structure, and the \bah{138} calculation tests whether it remains
useful when the photon-scattering force is intrinsically weaker.
This comparison is intended to assess the portability of direct CB-MOT loading
rather than to introduce a new force-calculation technique.
\subsection{\texorpdfstring{\baf{138}}{138BaF} benchmark}

We begin with \baf{138}, where laser slowing, red-detuned trapping, blue-detuned
compression, and direct CB-MOT loading have been studied experimentally \cite{
	chenStructureBranchingRatios2016,buRotationalHyperfineStructure2017,
	zhangLaserCoolingSlowing2022,zengThreeDimensionalMagnetoOpticalTrapping2024,
	zengDirectLoadingBaF2026}.
The main cooling transition is
\(\ket{X\Sigma,N=1}\rightarrow\ket{A\Pi_{1/2},+}\), with linewidth
\(\Gamma=2\pi\times\SI{2.84}{\mega\hertz}\) and wavelength
\(\lambda=\SI{860}{\nano\meter}\).
For a beam propagating along the quantization axis, \(\sigma^+\) light is taken
to drive \(\Delta m=+1\) transitions.

Figure~\ref{fig:baf138_force_map} compares the calculated axial forces for
representative red-MOT and CB-MOT parameters. Both configurations generate a
restoring force near the field zero, but their velocity structure is
qualitatively different.{In the red-detuned MOT, the decelerating
		force is confined to a comparatively narrow velocity range, and a
		low-velocity heating region appears in Fig.~\ref{fig:baf138_force_map}(c).}
This behavior is a common feature of type-II molecular MOTs, where Doppler
cooling competes with polarization-gradient heating
\cite{
	emileMagneticallyAssistedSisyphus1993,tarbuttMagnetoOpticalTrappingForces2015,
	devlinThreeDimensionalDopplerPolarizationGradient2016,
	devlinLaserCoolingMagnetoOptical2018,jarvisBlueDetunedMagnetoOpticalTrap2018}.
	{The CB-MOT instead produces a broad decelerating region associated
		with blue-detuned Sisyphus cooling in the moving conveyor-belt frame. As shown
		in Fig.~\ref{fig:baf138_force_map}(e), the low-velocity heating region is
		removed, consistent with the lower temperatures observed experimentally in the
		CB-MOT \cite{zengDirectLoadingBaF2026}.}


\begin{figure}[tbp]
	\centering
	\includegraphics[width=0.46\textwidth]{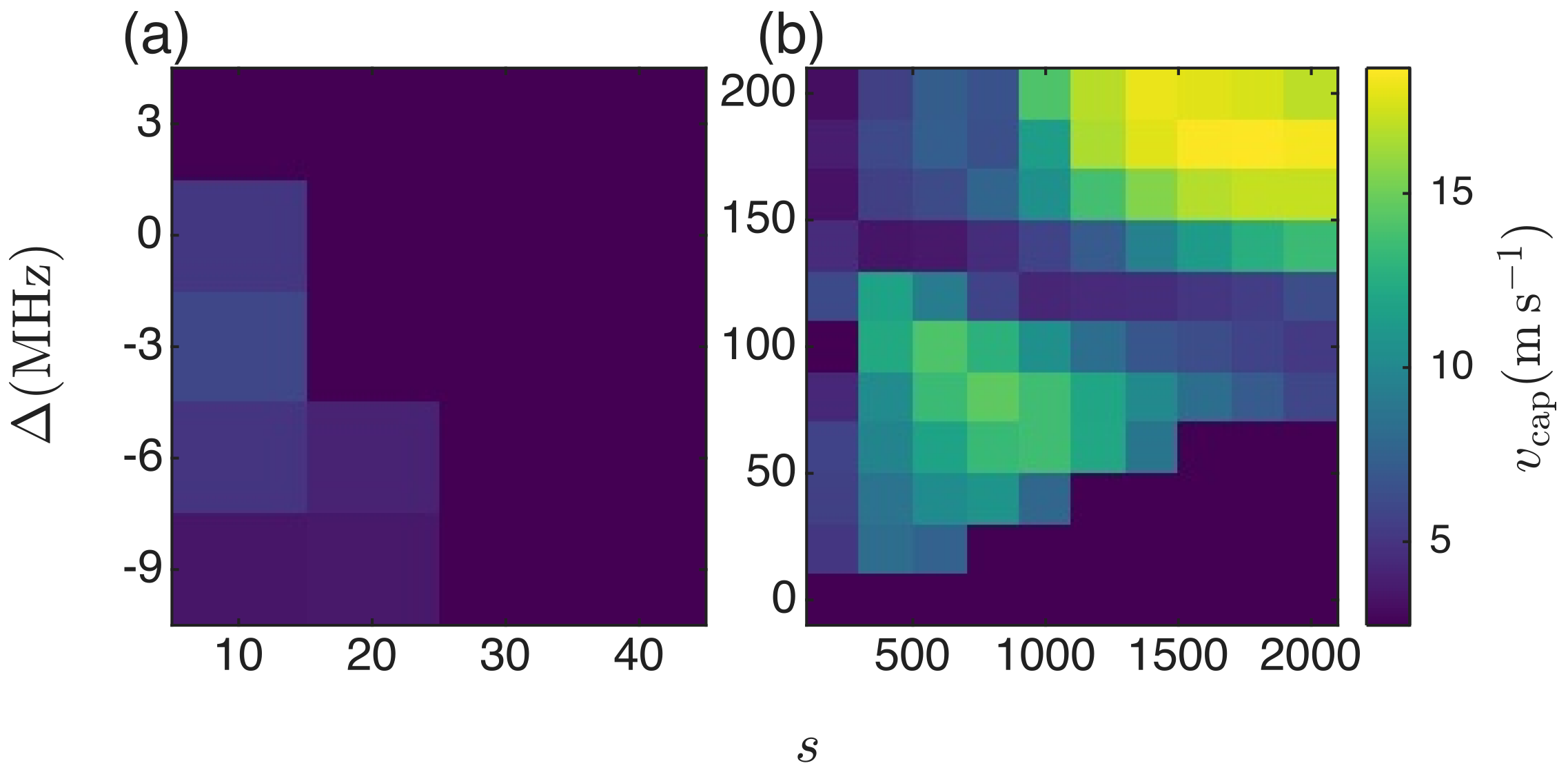}
	\caption{(a) Capture-velocity map with \baf{138} for dual-frequency red MOT.
		The frequency separations for the four laser frequencies are set the
		{same as in Fig.~\ref{fig:baf138_force_map}(a), and
		\(\Delta=\Delta_1\) is scanned.}
		(b) Capture-velocity map with \baf{138} for CB-MOT.
		The two-photon detunings are \(\left[\delta_{1}, \delta_{2} \right] =
		2\pi\times \left[-2, -3 \right]\,\si{\mega\hertz} \)\revise{red}{.}
	}.
	\label{fig:baf138_capture_phase}
\end{figure}

{The intensity dependence in Fig.~\ref{fig:baf138_force_map}
	separates damping from confinement. For the red MOT, increasing the laser
	intensity initially increases the available decelerating force, but it also
	enhances the low-velocity heating feature and can weaken or even reverse the
	restoring force. As shown in Fig.~\ref{fig:baf138_force_map}(d), the red MOT
	becomes anti-trapping at large saturation parameters. This explains why the
	red MOT must be operated in a relatively narrow intensity range: the cooling
	and trapping forces are ultimately limited by the radiative force scale
	\(\hbar k\Gamma\), which is especially small for heavy molecules with narrow
	transition linewidths.}

{The CB-MOT shows the opposite trend over the same range of
	parameters. The damping force grows and broadens with increasing saturation
	parameter, while the restoring force remains favorable over a larger magnetic
	field range, as shown in Figs.~\ref{fig:baf138_force_map}(e) and
	\ref{fig:baf138_force_map}(f). This behavior reflects the
	dipole-force-dominated nature of the conveyor-belt mechanism. Once molecules
	are cooled into one of the moving polarization gradients, the force is not set
	directly by the maximum scattering rate or by a single-photon recoil kick.
	Instead, larger optical power increases the AC Stark shifts and extends the
	velocity range over which moving-frame Sisyphus cooling is effective
	\cite{liConveyorBeltMagnetoOpticalTrapping2025,zengDirectLoadingBaF2026}.}

{The overlaid trajectories in Figs.~\ref{fig:baf138_force_map}(a)
	and \ref{fig:baf138_force_map}(b) illustrate how the local force map is
	converted into a capture criterion. A molecule entering the MOT samples a wide
	range of \(z\) and \(v\), and it is lost if the integrated deceleration is
	insufficient before it leaves the beam volume.}

{Figure~\ref{fig:baf138_capture_phase} shows the resulting capture
	velocity maps as a function of laser intensity and detuning for the red MOT
	and the CB-MOT. The two configurations occupy very different usable parameter
	space. The red MOT has nonzero capture velocity only in a narrow region where
	both damping and restoring forces remain favorable; the optimal intensity is
	around \(s\sim10\), and the detuning window is only a few megahertz,
	consistent with experimental observations.}

{In the CB-MOT, increasing laser intensity opens a broad
	high-\(\vc\) region. The weak dependence on the precise single-photon detuning
	is experimentally useful because it reduces sensitivity to laser-frequency
	drifts and calibration errors. The decrease near
	$\Delta \sim 2\pi\times 150 $~MHz occurs when additional hyperfine
	resonances begin to compete with the desired conveyor-belt force.
	{For the experimental CB-MOT condition \(s=200\), the calculated
	capture velocity is \(v_\mathrm{cap}\simeq 7\) m/s, which is larger than the
	typical red-MOT value of \(v_\mathrm{cap}\simeq 6\) m/s.} The large
	CB-MOT capture velocity is consistent with the direct-loading experiment of
	Ref.~\cite{zengDirectLoadingBaF2026}, where the loading efficiency increased
	with optical power and exceeded that of the red MOT.}

\subsection{\texorpdfstring{\baf{137}}{137BaF}: dense hyperfine structure}

{We next consider the fermionic isotope \baf{137}, which is a good
candidate for measurements of nuclear-spin-dependent parity violation and has
recently been transversely laser cooled
\cite{kogelLasercooled137mathrmBaFMolecules2025}. This isotope provides a
direct test of whether CB-MOT loading can be extended to molecules whose
internal structure makes a conventional red MOT difficult to optimize.} The
optical wavelength and excited-state linewidth are essentially the same as in
\baf{138}, but the
nonzero nuclear spin produces a much denser hyperfine structure. For a
conventional red-detuned MOT, this creates a practical difficulty: many
ground-state components must be addressed while maintaining the correct
detunings and polarizations for both damping and restoring forces.
{The CB-MOT is attractive in this case because the same moving-frame
Sisyphus mechanism can operate even when several hyperfine components are
included.}

\begin{figure}[tbp]
	\centering
	\includegraphics[width=0.40\textwidth]{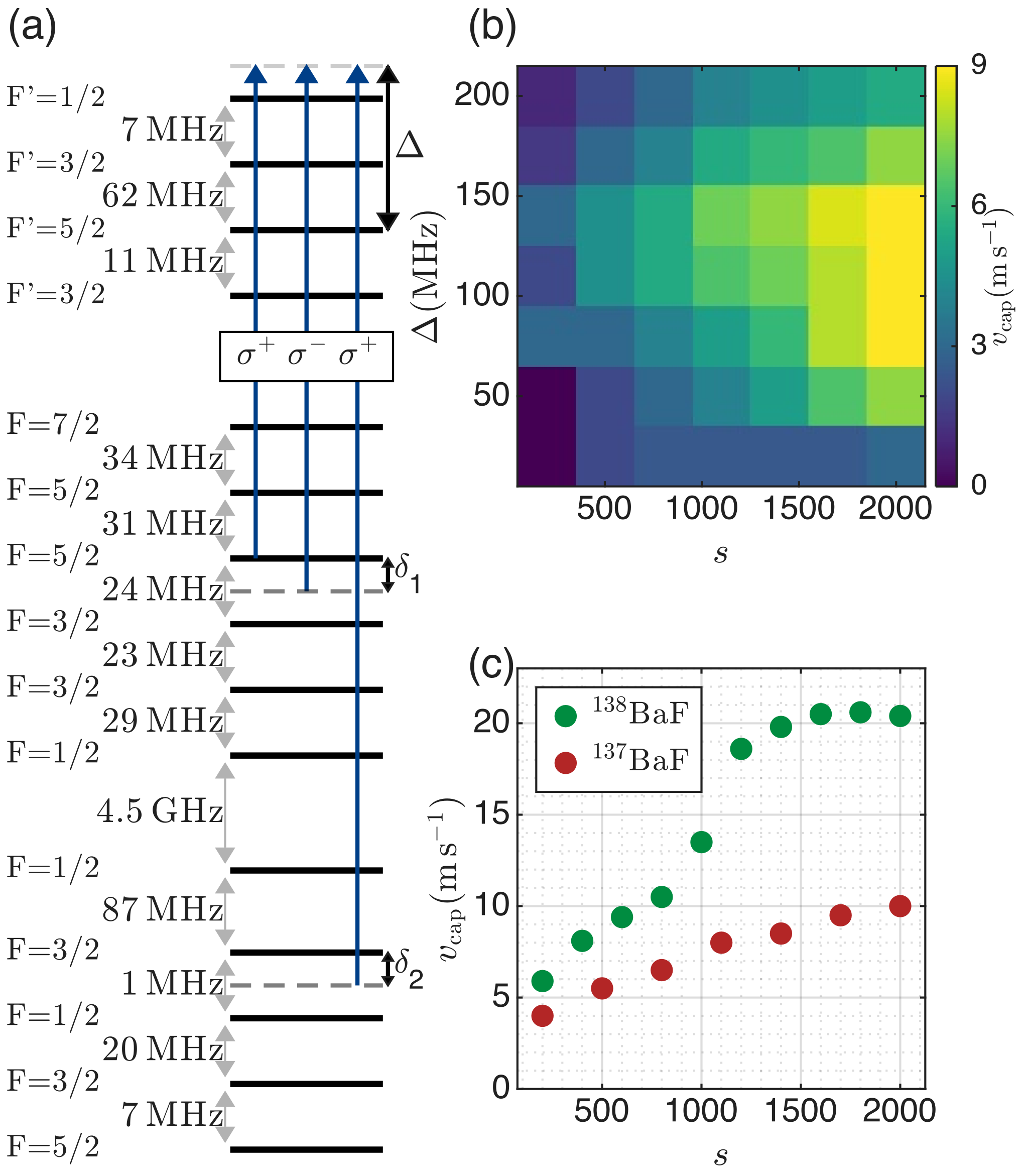}
	\caption{CB-MOT calculation for \baf{137}. (a) Laser-frequency scheme used to
		address the relevant hyperfine components. {The chosen
				{two-photon detunings are
				\([\delta_1, \delta_2]=2\pi\times[-2, -3]\) MHz.} (b)
				Capture velocity map for the \baf{137} CB-MOT. (c) Comparison of capture
				velocity versus saturation parameter for \baf{137} at $\Delta=2\pi\times 60$ MHz and \baf{138} at {$\Delta=2\pi\times 10$ MHz}.
			}}
	\label{fig:baf_isotope_capture}
\end{figure}
{Figure~\ref{fig:baf_isotope_capture} summarizes the \baf{137}
	calculation. The relevant hyperfine levels and the "1+2" laser scheme are shown
	in Fig.~\ref{fig:baf_isotope_capture}(a). The chosen three-frequency scheme
	addresses the dominant hyperfine components while preserving the
	conveyor-belt configuration; the detunings are defined in the figure.}

{Figure~\ref{fig:baf_isotope_capture}(b) shows the capture velocity
	versus saturation parameter \(s\) and single-photon detuning \(\Delta\). Each
	beam contains three frequency components with the same saturation parameter.
	A broad region of nonzero \(\vc\) appears at large \(s\). The weak dependence
	on \(\Delta\) persists over a detuning window of tens of megahertz, which is
	an important practical feature for realizing MOTs in molecules with dense
	hyperfine structure.}

{Figure~\ref{fig:baf_isotope_capture}(c) compares \(\vc\) for
	\baf{137} and \baf{138}. The capture velocity of \baf{137} is smaller, as
	expected from the larger number of coupled ground states and the corresponding
	reduction of optical coupling strength in each addressed component.
	Nevertheless, the qualitative CB-MOT scaling remains: increasing intensity
	expands the velocity range over which molecules can be slowed into the trap.
	At experimentally accessible saturation parameters
	\cite{zengDirectLoadingBaF2026}, {for example \(s=200\),} the calculation predicts a finite capture
	velocity {of about 4 m/s} for \baf{137}, suggesting that this isotope can be loaded with a
	CB-MOT even though a clean red-MOT implementation would be difficult to design
	and optimize.}

\subsection{\texorpdfstring{\bah{138}}{138BaH}: reduced radiative-force scale}

{The molecule \bah{138}, which has been transversely laser cooled
	\cite{mcnallyOpticalCyclingRadiative2020}, tests a different limitation.} Instead
of hyperfine complexity, the main challenge is the smaller radiative force. For
the transition considered here, \(\Gamma=2\pi\times\SI{1.2}{\mega\hertz}\) and
\(\lambda=\SI{1060}{\nano\meter}\). Both the narrower linewidth and the longer
wavelength reduce \(\hbar k\Gamma\), making photon-scattering-based slowing less
effective for a molecule of comparable mass. This is precisely the regime in
which a dipole-force-dominated loading mechanism should be most beneficial,
{so BaH provides a complementary test of the generality of direct
CB-MOT loading.}

\begin{figure}[tbp]
	\centering
	\includegraphics[width=0.40\textwidth]{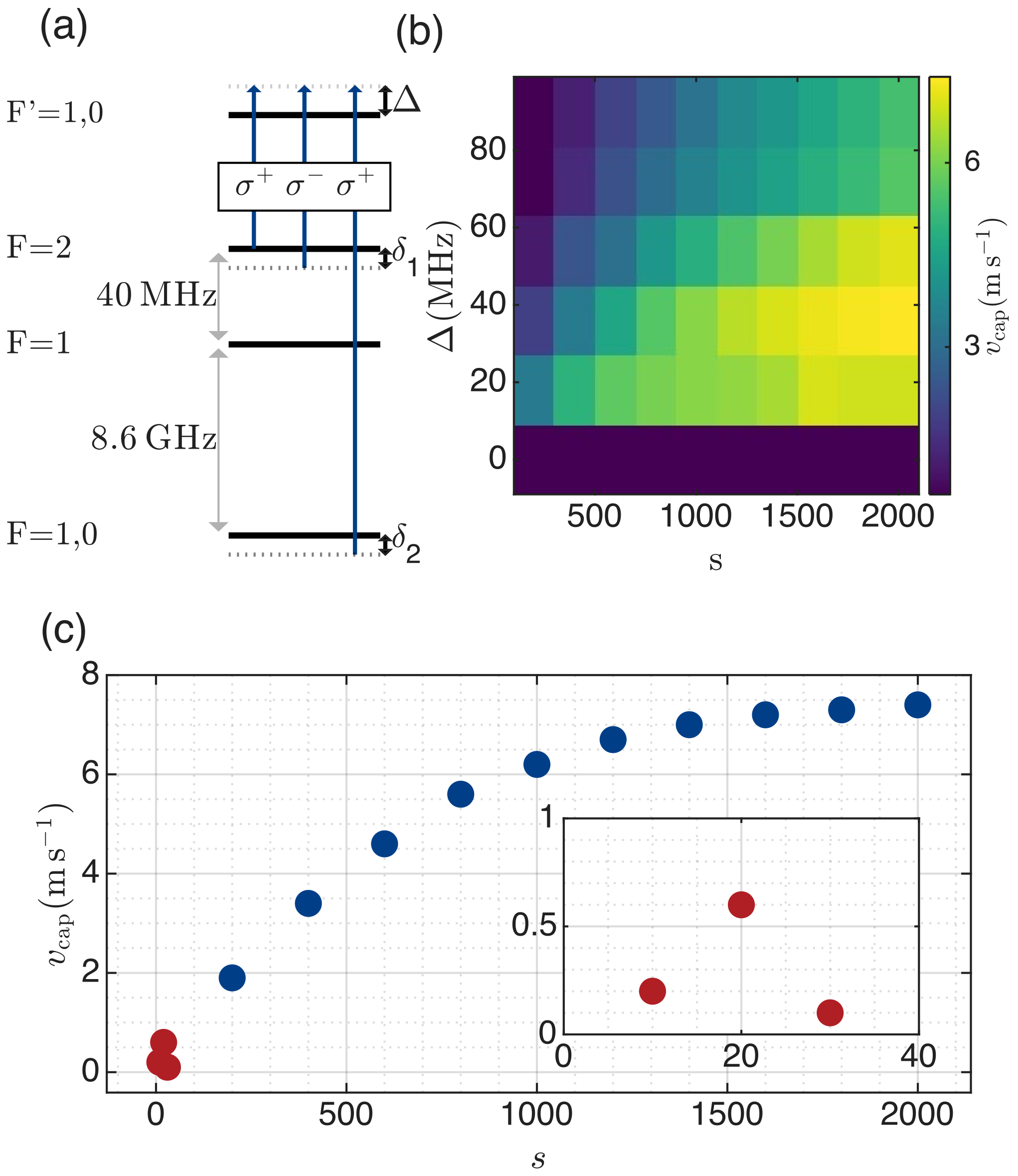}
	\caption{CB-MOT calculation for \bah{138}. (a) Laser scheme used in the CB-MOT. (b)
		Capture velocity map versus saturation parameter \(s\) and
		single-photon detuning \(\Delta\), with fixed two-photon detunings as defined
		in panel (a), {\([\delta_1, \delta_2]=2\pi\times[-2, -3]\) MHz.} (c) Capture velocity for the red MOT (red points)
		and the CB-MOT (blue points) at {$\Delta=2\pi\times 10$ MHz}.
	}
	\label{fig:bah138_capture}
\end{figure}

{The numerical results confirm that the CB-MOT can provide a much
	larger capture velocity even when the radiative-force scale is reduced.
	Figure~\ref{fig:bah138_capture}(a) shows the \bah{138} CB-MOT configuration,
	which uses the same "1+2" structure as in BaF. Figure~\ref{fig:bah138_capture}(b)
	shows the capture velocity map versus \(s\) and \(\Delta\). As in BaF, the
	capture velocity increases with laser intensity and is only weakly sensitive
	to the precise single-photon detuning over a broad range.}

{Figure~\ref{fig:bah138_capture}(c) compares the dual-frequency red
	MOT with the CB-MOT. The red-MOT capture velocity remains below
	1 m$/$s for the parameters considered here, smaller than
	the value estimated in Ref.~\cite{mcnallyOpticalCyclingRadiative2020}. This
	difference is expected because that work considered an AC-MOT configuration
	and used a multilevel rate-equation model, whereas the present calculation
	treats a DC-MOT using optical Bloch equations. {With such a small
	capture velocity, implementing a dual-frequency red MOT would be
	challenging.}

	On the other hand, the CB-MOT capture velocity is
	substantially larger and increases continuously with $s$. Since \bah{138} has a narrow excited-state linewidth
	{and a low saturation intensity, \(I_s=0.128\) mW/cm\(^2\),
	almost five times smaller than that of BaF, \(s=1000\) should be
	experimentally accessible. In this regime, the calculated CB-MOT capture
	velocity reaches a relatively large value of \(\sim 6\) m/s.} These results show that direct CB-MOT loading can remain useful even when the radiative force scale is reduced. More generally, it suggests that the conveyor-belt mechanism can benefit heavy laser-coolable species for which narrow linewidths
	and small recoil velocities make red-detuned MOT loading inefficient.}

\section{Conclusion}

We have used established optical-Bloch-equation force calculations and
classical trajectory propagation to evaluate the capture velocity of CB-MOTs in barium-containing molecules. For \baf{138}, the 	calculation explains the key observation of recent direct-loading 	experiments: the CB-MOT can reach a larger and more robust capture velocity than a conventional red MOT because its force is dominated by blue-detuned Sisyphus cooling in moving polarization gradients rather than by photon 	scattering alone. The red MOT is constrained by the need to balance damping
and restoring forces, whereas the CB-MOT maintains confinement while its useful velocity range expands with laser intensity.

Our simulation also shows that this direct-loading mechanism is not tied to the 	particular level structure of bosonic BaF. The \baf{137} calculation shows 	that a dense hyperfine spectrum reduces the attainable capture velocity but does not eliminate the broad high-intensity CB-MOT loading region. The \bah{138} calculation shows that even when the radiative-force scale is reduced by a narrower linewidth and longer wavelength, the CB-MOT can still provide a capture velocity on the few m$/$s scale over a broad range of parameters. These two extension tests support direct CB-MOT loading as a general strategy for heavy molecular species whose conventional red-detuned MOTs are limited by hyperfine complexity, weak scattering forces, or	technical sensitivity to detuning.

\begin{acknowledgments}
	We acknowledge the support from  the National Natural Science
	Foundation of China under Grant Nos. 92576201, and 12425408, the Innovation Program for Quantum Science and
	Technology under Grant No. 2024ZD0300601, the	National Key Research and Development Program of China under Grant
	No.2023YFA1406703 and No. 2022YFA1404203, the Zhejiang Provincial Applied Basic
	Research Program under Grant No. 2026C02A2005, and the Fundamental Research
	Funds for the Central Universities under Grant No. 2024FZZX02-01-02.
\end{acknowledgments}

\bibliographystyle{apsrev4-2}
\bibliography{references}

@article{aikawaCoherentTransferPhotoassociated2010,
  title = {Coherent {{Transfer}} of {{Photoassociated Molecules}} into the {{Rovibrational Ground State}}},
  author = {Aikawa, K. and Akamatsu, D. and Hayashi, M. and Oasa, K. and Kobayashi, J. and Naidon, P. and Kishimoto, T. and Ueda, M. and Inouye, S.},
  year = 2010,
  journal = {Physical Review Letters},
  volume = {105},
  pages = {203001},
  doi = {10.1103/physrevlett.105.203001}
}

@article{andereggLaserCoolingOptically2018,
  title = {Laser {{Cooling}} of {{Optically Trapped Molecules}}},
  author = {Anderegg, L. and Augenbraun, B. L. and Bao, Y. and Burchesky, S. and Cheuk, L. W. and Ketterle, W. and Ni, K.-K. and Doyle, J. M.},
  year = 2018,
  journal = {Nature Physics},
  volume = {14},
  pages = {890},
  doi = {10.1038/s41567-018-0191-z}
}

@article{Dai2026MOT,
  title = {Magneto-Optical Trapping of a Metal Hydride Molecule},
  author = {Dai, Jinyu and Riley, Benjamin and Sun, Qi and Mitra, Debayan and Zelevinsky, Tanya},
  journal = {Phys. Rev. Lett.},
  volume = {136},
  issue = {23},
  pages = {233403},
  numpages = {7},
  year = {2026},
  month = {Jun},
  publisher = {American Physical Society},
  doi = {10.1103/xy6y-kyhc},
  url = {https://link.aps.org/doi/10.1103/xy6y-kyhc}
}

@article{Padilla2025AlFMOT,
  title = {Magneto-Optical Trapping of Aluminum Monofluoride},
  author = {Padilla-Castillo, J. E. and Cai, J. and Agarwal, P. and Kukreja, P. and Thomas, R. and Sartakov, B. G. and Truppe, S. and Meijer, G. and Wright, S. C.},
  journal = {Phys. Rev. Lett.},
  volume = {135},
  issue = {24},
  pages = {243401},
  numpages = {7},
  year = {2025},
  month = {Dec},
  publisher = {American Physical Society},
  doi = {10.1103/ksnd-9fyf},
  url = {https://link.aps.org/doi/10.1103/ksnd-9fyf}
}

@article{andereggOpticalTweezerArray2019,
  title = {An {{Optical Tweezer Array}} of {{Ultracold Molecules}}},
  author = {Anderegg, L. and Cheuk, L. W. and Bao, Y. and Burchesky, S. and Ketterle, W. and Ni, K.-K. and Doyle, J. M.},
  year = 2019,
  journal = {Science},
  volume = {365},
  pages = {1156},
  doi = {10.1126/science.aax1265}
}

@article{andereggRadioFrequencyMagnetoOptical2017,
  title = {Radio {{Frequency Magneto-Optical Trapping}} of {{CaF}} with {{High Density}}},
  author = {Anderegg, L. and Augenbraun, B. L. and Chae, E. and Hemmerling, B. and Hutzler, N. R. and Ravi, A. and Collopy, A. and Ye, J. and Ketterle, W. and Doyle, J. M.},
  year = 2017,
  journal = {Physical Review Letters},
  volume = {119},
  pages = {103201},
  doi = {10.1103/physrevlett.119.103201}
}

@article{baoDipolarSpinExchangeEntanglement2023,
  title = {Dipolar {{Spin-Exchange}} and {{Entanglement}} between {{Molecules}} in an {{Optical Tweezer Array}}},
  author = {Bao, Y. and Yu, S. S. and Anderegg, L. and Chae, E. and Ketterle, W. and Ni, K.-K. and Doyle, J. M.},
  year = 2023,
  journal = {Science},
  volume = {382},
  pages = {1138},
  doi = {10.1126/science.adf8999}
}

@article{barryMagnetoOpticalTrappingDiatomic2014,
  title = {Magneto-{{Optical Trapping}} of a {{Diatomic Molecule}}},
  author = {Barry, J. F. and McCarron, D. J. and Norrgard, E. B. and Steinecker, M. H. and DeMille, D.},
  year = 2014,
  journal = {Nature},
  volume = {512},
  pages = {286},
  doi = {10.1038/nature13634}
}

@article{bauseLaserCoolingCandidate2025,
  title = {Laser {{Cooling Candidate BaOH}} for {{Precision Measurements}}},
  author = {Bause, R. and Balasubramanian, N. and Fikkers, T. and Prinsen, E. H. and Steinebach, K. and Jadbabaie, A. and Hutzler, N. R. and Aucar, I. A. and Pasteka, L. F. and Borschevsky, A. and Hoekstra, S.},
  year = 2025,
  journal = {Physical Review A},
  volume = {111},
  pages = {062815},
  doi = {10.1103/PhysRevA.111.062815}
}

@article{burauBlueDetunedMagnetoOpticalTrap2023,
  title = {Blue-{{Detuned Magneto-Optical Trap}} of {{Molecules}}},
  author = {Burau, J. J. and Aggarwal, P. and Mehling, K. and Ye, J.},
  year = 2023,
  journal = {Physical Review Letters},
  volume = {130},
  pages = {193401},
  doi = {10.1103/physrevlett.130.193401}
}

@article{buRotationalHyperfineStructure2017,
  title = {Rotational and {{Hyperfine Structure}} of {{BaF}} for {{Laser Cooling}}},
  author = {Bu, W. and Chen, T. and Lv, G. and Yan, B.},
  year = 2017,
  journal = {Physical Review A},
  volume = {95},
  pages = {032701},
  doi = {10.1103/physreva.96.053401}
}

@article{carrColdUltracoldMolecules2009,
  title = {Cold and Ultracold Molecules: Science, Technology and Applications},
  author = {Carr, Lincoln D. and DeMille, David and Krems, Roman V. and Ye, Jun},
  year = 2009,
  month = may,
  journal = {New J. Phys.},
  volume = {11},
  pages = {055049},
  issn = {1367-2630},
  doi = {10.1088/1367-2630/11/5/055049}
}

@article{chenStructureBranchingRatios2016,
  title = {Structure, {{Branching Ratios}}, and a {{Laser-Cooling Scheme}} for the {{BaF}} 138 {{Molecule}}},
  author = {Chen, Tao and Bu, Wenhao and Yan, Bo},
  year = 2016,
  month = dec,
  journal = {Physical Review A},
  volume = {94},
  number = {6},
  pages = {063415},
  issn = {2469-9926, 2469-9934},
  doi = {10.1103/PhysRevA.94.063415},
  urldate = {2023-03-28},
  lccn = {2}
}

@article{collopy3DMagnetoOpticalTrap2018,
  title = {{{3D Magneto-Optical Trap}} of {{Yttrium Monoxide}}},
  author = {Collopy, A. L. and Ding, S. and Wu, Y. and Finneran, I. A. and Anderegg, L. and Augenbraun, B. L. and Doyle, J. M. and Ye, J.},
  year = 2018,
  journal = {Physical Review Letters},
  volume = {121},
  pages = {213201},
  doi = {10.1103/physrevlett.121.213201}
}

@article{cornishQuantumComputationQuantum2024,
  title = {Quantum Computation and Quantum Simulation with Ultracold Molecules},
  author = {Cornish, Simon L and Tarbutt, Michael R and Hazzard, Kaden RA},
  year = 2024,
  journal = {Nature Physics},
  volume = {20},
  number = {5},
  pages = {730--740},
  publisher = {Nature Publishing Group UK London},
  doi = {10.1038/s41567-024-02453-9}
}

@article{demilleQuantumSensingMetrology2024,
  title = {Quantum Sensing and Metrology for Fundamental Physics with Molecules},
  author = {DeMille, David and Hutzler, Nicholas R and Rey, Ana Maria and Zelevinsky, Tanya},
  year = 2024,
  journal = {Nature Physics},
  volume = {20},
  number = {5},
  pages = {741--749},
  publisher = {Nature Publishing Group UK London},
  doi = {10.1038/s41567-024-02499-9}
}

@article{devlinLaserCoolingMagnetoOptical2018,
  title = {Laser {{Cooling}} and {{Magneto-Optical Trapping}} of {{Molecules Analyzed Using Optical Bloch Equations}} and the {{Fokker-Planck-Kramers Equation}}},
  author = {Devlin, J. A. and Tarbutt, M. R.},
  year = 2018,
  month = dec,
  journal = {Physical Review A},
  volume = {98},
  number = {6},
  pages = {063415},
  publisher = {American Physical Society},
  doi = {10.1103/PhysRevA.98.063415},
  urldate = {2024-03-11},
  lccn = {2}
}

@article{devlinThreeDimensionalDopplerPolarizationGradient2016,
  title = {Three-{{Dimensional Doppler}}, {{Polarization-Gradient}} and {{Magneto-Optical Forces}} for {{Atoms}} and {{Molecules}} with {{Dark States}}},
  author = {Devlin, J. A. and Tarbutt, M. R.},
  year = 2016,
  journal = {New Journal of Physics},
  volume = {18},
  pages = {123017},
  doi = {10.1088/1367-2630/18/12/123017}
}

@article{dirosaLasercoolingMoleculesConcept2004,
  title = {Laser-Cooling Molecules: {{Concept}}, Candidates, and Supporting Hyperfine-Resolved Measurements of Rotational Lines in the {{AX}} (0, 0) Band of {{CaH}}},
  author = {Di Rosa, {\relax MD}},
  year = 2004,
  journal = {The European Physical Journal D-Atomic, Molecular, Optical and Plasma Physics},
  volume = {31},
  number = {2},
  pages = {395--402},
  publisher = {Springer},
  doi = {10.1140/epjd/e2004-00167-2}
}

@article{emileMagneticallyAssistedSisyphus1993,
  title = {Magnetically {{Assisted Sisyphus Effect}}},
  author = {Emile, O. and Kaiser, R. and Gerz, C. and Wallis, H. and Aspect, A. and {Cohen-Tannoudji}, C.},
  year = 1993,
  journal = {Journal de Physique II},
  volume = {3},
  pages = {1709},
  doi = {10.1051/jp2:1993226}
}

@article{hallasHighCompressionBlueDetunedMagnetoOptical2026,
  title = {High-{{Compression Blue-Detuned Magneto-Optical Trap}} of {{Polyatomic Molecules}}},
  author = {Hallas, C. and Li, G. K. and Vilas, N. B. and Robichaud, P. and Anderegg, L. and Doyle, J. M.},
  year = 2026,
  journal = {Physical Review Letters},
  doi = {10.1103/w9qc-rczf}
}

@article{hollandOnDemandEntanglementMolecules2023,
  title = {On-{{Demand Entanglement}} of {{Molecules}} in a {{Reconfigurable Optical Tweezer Array}}},
  author = {Holland, C. M. and Lu, Y. and Cheuk, L. W.},
  year = 2023,
  journal = {Science},
  volume = {382},
  pages = {1143},
  doi = {10.1126/science.adf4272}
}

@article{jarvisBlueDetunedMagnetoOpticalTrap2018,
  title = {Blue-{{Detuned Magneto-Optical Trap}}},
  author = {Jarvis, K. N. and Devlin, J. A. and Wall, T. E. and Sauer, B. E. and Tarbutt, M. R.},
  year = 2018,
  month = feb,
  journal = {Physical Review Letters},
  volume = {120},
  number = {8},
  pages = {083201},
  issn = {0031-9007, 1079-7114},
  doi = {10.1103/PhysRevLett.120.083201},
  urldate = {2024-02-29},
  lccn = {1}
}

@article{jorapurHighDensityLoading2024,
  title = {High {{Density Loading}} and {{Collisional Loss}} of {{Laser-Cooled Molecules}} in an {{Optical Trap}}},
  author = {Jorapur, V. and Langin, T. K. and Wang, Q. and Zheng, G. and DeMille, D.},
  year = 2024,
  journal = {Physical Review Letters},
  volume = {132},
  pages = {163403},
  doi = {10.1103/physrevlett.132.163403}
}

@article{karmanUltracoldChemistryTestbed2024,
  title = {Ultracold Chemistry as a Testbed for Few-Body Physics},
  author = {Karman, Tijs and Tomza, Micha{\l} and {P{\'e}rez-R{\'i}os}, Jes{\'u}s},
  year = 2024,
  journal = {Nature Physics},
  volume = {20},
  number = {5},
  pages = {722--729},
  publisher = {Nature Publishing Group UK London},
  doi = {10.1038/s41567-024-02467-3}
}

@article{kogelLasercooled137mathrmBaFMolecules2025,
  title = {Laser-Cooled {$^{137}\mathrm{BaF}$} Molecules for Measuring Nuclear-Spin-Dependent Parity Violation},
  author = {Kogel, Felix and Garg, Tatsam and Rockenh{\"a}user, Marian and Langen, Tim},
  year = 2025,
  month = may,
  journal = {Phys. Rev. Res.},
  volume = {7},
  number = {2},
  pages = {L022041},
  publisher = {American Physical Society},
  doi = {10.1103/PhysRevResearch.7.L022041}
}

@article{langenQuantumStateManipulation2024,
  title = {Quantum State Manipulation and Cooling of Ultracold Molecules},
  author = {Langen, Tim and Valtolina, Giacomo and Wang, Dajun and Ye, Jun},
  year = 2024,
  journal = {Nature Physics},
  volume = {20},
  number = {5},
  pages = {702--712},
  publisher = {Nature Publishing Group UK London},
  doi = {10.1038/s41567-024-02423-1}
}

@article{langinTheoryMolecularMagnetoOptical2023,
  title = {Theory of the {{Molecular Magneto-Optical Trap}}},
  author = {Langin, T. K. and DeMille, D.},
  year = 2023,
  journal = {New Journal of Physics},
  volume = {25},
  pages = {043005},
  doi = {10.1088/1367-2630/acc34d}
}

@article{lasnerMagnetoOpticalTrappingHeavy2025,
  title = {Magneto-{{Optical Trapping}} of a {{Heavy Polyatomic Molecule}} for {{Precision Measurement}}},
  author = {Lasner, Z. D. and Frenett, A. and Sawaoka, H. and Anderegg, L. and Augenbraun, B. L. and Lampson, H. and Li, M. and Lunstad, A. and Mango, J. and Nasir, A. and Ono, T. and Sakamoto, T. and Doyle, J. M.},
  year = 2025,
  journal = {Physical Review Letters},
  volume = {134},
  pages = {083401},
  doi = {10.1103/physrevlett.134.083401}
}

@article{liBlueDetunedMagnetoOpticalTrap2024,
  title = {Blue-{{Detuned Magneto-Optical Trap}} of {{CaF Molecules}}},
  author = {Li, S. J. and Holland, C. M. and Lu, Y. and Cheuk, L. W.},
  year = 2024,
  journal = {Physical Review Letters},
  volume = {132},
  pages = {233402},
  doi = {10.1103/physrevlett.132.233402}
}

@article{liConveyorBeltMagnetoOpticalTrapping2025,
  title = {Conveyor-{{Belt Magneto-Optical Trapping}} of {{Molecules}}},
  author = {Li, Grace K and Hallas, Christian and Doyle, John M},
  year = 2025,
  month = apr,
  journal = {New Journal of Physics},
  volume = {27},
  number = {4},
  pages = {043002},
  issn = {1367-2630},
  doi = {10.1088/1367-2630/adc032},
  urldate = {2025-04-03}
}

@article{lyuTrappingCoolingMechanisms2026,
  title = {Trapping and Cooling Mechanisms in Blue-Detuned Magneto-Optical Traps of Molecules},
  author = {Lyu, Qinshu and Tarbutt, M. R.},
  year = 2026,
  month = jun,
  journal = {Phys. Rev. Res.},
  volume = {8},
  number = {2},
  pages = {023259},
  publisher = {American Physical Society},
  doi = {10.1103/ctj6-6hg9}
}

@article{mcnallyOpticalCyclingRadiative2020,
  title = {Optical {{Cycling}}, {{Radiative Deflection}}, and {{Laser Cooling}} of {{Barium Monohydride}}},
  author = {McNally, R. L. and Kozyryev, I. and {Vazquez-Carson}, S. and Wenz, K. and Wang, T. and Zelevinsky, T.},
  year = 2020,
  journal = {New Journal of Physics},
  volume = {22},
  pages = {083047},
  doi = {10.1088/1367-2630/aba3e9}
}

@article{niHighPhaseSpaceDensityGas2008,
  title = {A {{High Phase-Space-Density Gas}} of {{Polar Molecules}}},
  author = {Ni, K.-K. and Ospelkaus, S. and {de Miranda}, M. H. G. and Peer, A. and Neyenhuis, B. and Zirbel, J. J. and Kotochigova, S. and Julienne, P. S. and Jin, D. S. and Ye, J.},
  year = 2008,
  journal = {Science},
  volume = {322},
  pages = {231},
  doi = {10.1126/science.1163861}
}

@article{shumanLaserCoolingDiatomic2010,
  title = {Laser Cooling of a Diatomic Molecule},
  author = {Shuman, E. S. and Barry, J. F. and DeMille, D.},
  year = 2010,
  month = oct,
  journal = {Nature},
  volume = {467},
  number = {7317},
  pages = {820--823},
  issn = {0028-0836},
  doi = {10.1038/nature09443}
}

@article{stuhlMagnetoOpticalTrapPolar2008,
  title = {Magneto-{{Optical Trap}} for {{Polar Molecules}}},
  author = {Stuhl, B. K. and Sawyer, B. C. and Wang, D. and Ye, J.},
  year = 2008,
  journal = {Physical Review Letters},
  volume = {101},
  pages = {243002},
  doi = {10.1103/physrevlett.101.243002}
}

@article{tarbuttMagnetoOpticalTrappingForces2015,
  title = {Magneto-{{Optical Trapping Forces}} for {{Atoms}} and {{Molecules}} with {{Complex Level Structures}}},
  author = {Tarbutt, M. R.},
  year = 2015,
  journal = {New Journal of Physics},
  volume = {17},
  pages = {015007},
  doi = {10.1088/1367-2630/17/1/015007}
}

@article{truppeMoleculesCooledDoppler2017,
  title = {Molecules {{Cooled Below}} the {{Doppler Limit}}},
  author = {Truppe, S. and Williams, H. J. and Hambach, M. and Caldwell, L. and Fitch, N. J. and Hinds, E. A. and Sauer, B. E. and Tarbutt, M. R.},
  year = 2017,
  journal = {Nature Physics},
  volume = {13},
  pages = {1173},
  doi = {10.1038/nphys4241}
}

@article{vilasMagnetoOpticalTrappingSubDoppler2022,
  title = {Magneto-{{Optical Trapping}} and {{Sub-Doppler Cooling}} of a {{Polyatomic Molecule}}},
  author = {Vilas, N. B. and Hallas, C. and Anderegg, L. and Robichaud, P. and Winnicki, A. and Mitra, D. and Doyle, J. M.},
  year = 2022,
  journal = {Nature},
  volume = {606},
  pages = {70},
  doi = {10.1038/s41586-022-04620-5}
}

@article{vilasOpticalTweezerArray2024,
  title = {An {{Optical Tweezer Array}} of {{Ultracold Polyatomic Molecules}}},
  author = {Vilas, N. B. and Robichaud, P. and Hallas, C. and Li, G. K. and Anderegg, L. and Doyle, J. M.},
  year = 2024,
  journal = {Nature},
  volume = {628},
  pages = {282},
  doi = {10.1038/s41586-024-07199-1}
}

@article{xuNumericalStudyGrayMolasses2023,
  title = {Numerical {{Study}} of {{Gray-Molasses Cooling}} of {{Molecules}}},
  author = {Xu, S. and Li, R. and Xia, Y. and Siercke, M. and Ospelkaus, S.},
  year = 2023,
  journal = {Physical Review A},
  volume = {108},
  pages = {033102},
  doi = {10.1103/PhysRevA.108.033102}
}

@article{yanObservationDipolarSpinexchange2013,
  title = {Observation of Dipolar Spin-Exchange Interactions with Lattice-Confined Polar Molecules},
  author = {Yan, Bo and Moses, Steven A and Gadway, Bryce and Covey, Jacob P and Hazzard, Kaden RA and Rey, Ana Maria and Jin, Deborah S and Ye, Jun},
  year = 2013,
  journal = {Nature},
  volume = {501},
  number = {7468},
  pages = {521--525},
  publisher = {Nature Publishing Group UK London},
  doi = {10.1038/nature12483}
}

@article{yuConveyorBeltMagnetoOpticalTrap2026,
  title = {A {{Conveyor-Belt Magneto-Optical Trap}} of {{CaF}}},
  author = {Yu, Scarlett S. and You, Jiaqi and Bao, Yicheng and Anderegg, Lo{\"i}c and Hallas, Christian and Li, Grace K. and Lim, Dongkyu and Chae, Eunmi and Ketterle, Wolfgang and Ni, Kang-Kuen and Doyle, John M.},
  year = 2026,
  month = jan,
  journal = {Nature Communications},
  publisher = {arXiv},
  issn = {2041-1723},
  doi = {10.1038/s41467-025-67944-6},
  urldate = {2026-01-27}
}

@article{zengDirectLoadingBaF2026,
  title = {Direct {{Loading}} of {{BaF Molecules}} with a {{Conveyor-Belt Magneto-optical Trap}}},
  author = {Zeng, Zixuan and Yang, Shoukang and Deng, Shuhua and Yan, Bo},
  year = 2026,
  month = feb,
  journal = {Physical Review Letters},
  volume = {136},
  number = {7},
  pages = {073402},
  issn = {0031-9007, 1079-7114},
  doi = {10.1103/pd77-s994},
  urldate = {2026-05-29}
}

@article{zengThreeDimensionalMagnetoOpticalTrapping2024,
  title = {Three-{{Dimensional Magneto-Optical Trapping}} of {{BaF Molecules}}},
  author = {Zeng, Z. and Deng, S. and Yang, S. and Yan, B.},
  year = 2024,
  journal = {Physical Review Letters},
  volume = {133},
  pages = {143404},
  doi = {10.1103/physrevlett.133.143404}
}

@article{zhangLaserCoolingSlowing2022,
  title = {Laser {{Cooling}} and {{Slowing}} of {{BaF Molecules}}},
  author = {Zhang, Y. and Zeng, Z. and Liang, Q. and Bu, W. and Yan, B.},
  year = 2022,
  journal = {Physical Review A},
  volume = {105},
  pages = {033307},
  doi = {10.1103/PhysRevA.105.033307}
}

@article{zhangOpticalTweezerArray2022,
  title = {An {{Optical Tweezer Array}} of {{Ground-State Polar Molecules}}},
  author = {Zhang, J. T. and Picard, L. R. B. and Cairncross, W. B. and Wang, K. and Yu, Y. and Fang, F. and Ni, K.-K.},
  year = 2022,
  journal = {Quantum Science and Technology},
  volume = {7},
  pages = {035006},
  doi = {10.1088/2058-9565/ac676c}
}

\end{document}